\documentclass[twocolumn,showpacs,amsmath,amssymb]{revtex4}
\usepackage[usenames]{color}
\usepackage{bm}
\usepackage{graphicx}
\def\vg#1{\boldsymbol #1}
\def\v#1{\bold #1}
\def\dfrac#1#2{{\displaystyle\frac{#1}{#2}}}
\def\H{{\cal H}}
\def\HA{{\H^{A}}}
\def\HB{{\H^{B}}}
\def\M0{M_{\rm 0}}
\def\Msat{M_{\rm s}}
\def\JeffA{J^{A}_{\rm eff}}
\def\DeffA{D^{A}_{\rm eff}}
\def\JeffB{J^{B}_{\rm eff}}
\def\DeffB{D^{B}_{\rm eff}}

\newcommand{\ket}[1]{\mid\!{#1}\,\rangle}

\newcommand{\aver}[1]{\langle {#1}\rangle}

\begin{document}

\title{Frustration Induced Quantum Phases in Mixed Spin Chain with Frustrated Side Chains}

\author{Kazuo Hida$^{1}$ and Ken'ichi Takano$^{2}$ }

\affiliation{
$^{1}$Division of Material Science, 
Graduate School of Science and Engineering, \\ 
Saitama University, 
Saitama, Saitama, 338-8570, Japan \\
$^{2}$Toyota Technological Institute, 
Tenpaku-ku, Nagoya 468-8511, Japan }

\date{\today}

\begin{abstract}
A mixed Heisenberg spin chain with frustrated side chains is investigated by numerical and perturbational calculations.  
A frustration-induced quantum  partially polarized ferrimagnetic phase 
and a nonmagnetic spin quadrupolar phase are found 
adjacent to the conventional Lieb-Mattis type ferrimagnetic phase or the nonmagnetic singlet cluster solid  phases. 
The  partially polarized ferrimagnetic phase has  an incommensurate spin structure. 
{Similar structures are commonly found in other frustration-induced  partially polarized ferrimagnetic phases.}
Numerical results also suggest a series of almost critical nonmagnetic ground states in a highly frustrated regime if the side chain spins weakly couple to the main chain. 
\end{abstract}    


\maketitle

\section{Introduction} 

The interplay of frustration and quantum fluctuation has been extensively studied in a variety of low-dimensional quantum magnets. Even in one-dimensional cases, various  exotic quantum phenomena such as spontaneous dimerization,~\cite{mg} 1/3-plateau {with spontaneous trimerization},~\cite{oku} and transition between quantum and classical plateaus~\cite{ha} are reported. On the other hand, the mixed quantum spin chains {also} have a variety of ground states ranging from quantum ferrimagnetism~\cite{kura} to spin gap phases~\cite{kt1,takano,koga,th}. 

Recently, it has been reported that frustration induces 
a partially polarized ferrimagnetic  (PPF) phase ~\cite{ir,ym,kh,ncf} 
 in addition to 
 a conventional Lieb-Mattis type ferrimagnetic (LMF) phase. 
The  PPF phase appear 
  when  both frustration and  quantum fluctuations are fairly strong. 
It is  an interesting  issue how general such a ferrimagnetism is. 
Hence it is important to investigate the features of 
 PPF phases in various spin systems. 
We  are then motivated to find a  PPF phase 
 in  other 
 models and investigate  them 
 in detail.

We have introduced a spin chain with side chains
in a previous paper~\cite{th}. 
This spin chain has frustration owing to the interaction 
among spins in the main chain and  those  
of side chains. 
The  frustration  varies 
 in strength and in feature 
with  the variation of 
parameters  
in the model. 
Since we focused on the
spin-gap phases in ~\cite{th}, 
we have investigated  the parameter regimes where frustration 
is not strong enough to destroy the spin-gap phases. 
We then found two spin-gap phases and explained them 
by singlet cluster solid (SCS) pictures.

In the present work, we examine this model 
in a highly frustrated  regime. 
This regime, in which the frustration plays a central role, is of interest in its own right, 
since the model exhibits features very different from those 
in  the weak frustration regimes. We actually find  clear numerical evidences  not only for the above mentioned fascinating PPF phase but also for the spin quadrupolar (QP) phase~\cite{uimin,lai,su,fs1,fs2,ik,lauchli}. These phases are totally different from the conventional phases such as spin gap phases and the LMF phase which can be realized even in the unfrustrated case. They will be investigated in detail in the present paper. Also numerical data are obtained which suggest the possible existence of an exotic almost critical nonmagnetic ground state in the regime where the couplings between the side chain and main chain spins are weak but strongly frustrated.

The transition from a ferrimagnetic phase to a nonmagnetic phase in the present model takes place because the quantum fluctuation in the side-chains  destroys the ferrimagnetic long range order in the main chain. 
The mechanism of quantum destruction of ferromagnetism 
and ferrimagnetism has been less  studied than 
that of antiferromagnetism; 
the latter has been extensively studied in relation with 
the high-$T_{\rm c}$ superconductivity. Recently, however, experiments have been reported on the nonmagnetic ground states in  one- and two-dimensional materials~\cite{hase,kage} with ferromagnetic nearest neighbour and antiferromagnetic next nearest neighbour couplings. Theoretical investigation has also been carried out for corresponding models~\cite{hamada,tonehara,iq,momoi}. 

In  an unfrustrated ferrimagnet, the spontaneous magnetization is uniquely determined by the Lieb-Mattis theorem~\cite{lm}.  This type of quantum ferrimagnetism  has been investigated in detail~\cite{kura}.   {As far as the frustration is weak, the spontaneous magnetization remains locked to this value~\cite{irs}. This phase is the LMF phase~\cite{ym}. }
 The spontaneous magnetization in this phase is a simple fraction of the saturated magnetization. In contrast, the spontaneous magnetization in the  PPF phase continuously varies with the parameter characterizing the strength of frustration and is not a simple fraction of the saturated magnetization. The  PPF phase appears between the LMF  phase and  the nonmagnetic spin gap phases.  {This type of phase is first predicted in the pioneering work of Sachev and Senthil~\cite{ss} in the quantum rotor model. Bartosch, Kollar and  Kopietz~\cite{bkk} proposed a possibility of ferromagnetic Luttinger liquid in an itinerant one-dimensional Fermi system. The first explicit example of quantum  PPF phase induced by frustration in one-dimensional quantum spin systems was proposed by  Ivanov and Richter~\cite{ir} in a frustrated mixed spin ladder. Similar phases are also found by  Yoshikawa and Miyashita~\cite{ym} in a uniform spin chain and by one of the present authors in a trimerized zigzag chain~\cite{kh,ncf}}. 
We propose another example of the  PPF phase in the present mixed spin chain.  {The present example is substantially different from previous ones, because it is  accompanied by the destruction of the ferrimagnetic order in the main chain by the frustrated coupling to the quantum fluctuation in the side chains.}

The QP phase is well known for the spin-1 bilinear-biquadratic chain between the Haldane and ferromagnetic phases~\cite{uimin,lai,su,fs1,fs2,ik,lauchli}. The exact Bethe ansatz solution is available if the coefficients of the bilinear and  the biquadratic terms coincide with each other~\cite{uimin,lai,su}. Recently, similar phases are found in the frustrated two-dimensional Heisenberg model with ferromagnetic nearest neighbour interaction and antiferromagnetic next nearest neighbour interaction~\cite{momoi}. In this paper we explicitly show that our model reduces to the bilinear-biquadratic chain in appropriate limiting cases. It is also argued that the QP phase should appear  in a wide class of complex spin models between ferrimagnetic  and  spin gap phases.

This paper is organized as follows. In the next section, the model Hamiltonian is presented. Various limiting cases are discussed using the perturbational approximation from the strong coupling limit in  {Sec.} III. The numerically obtained ground state phases are explained in  {Sec.} IV. The properties of  PPF phase are described in detail in  {Sec.} V.   {Section} VI is devoted to summary and discussion.

\section{Hamiltonian}
\begin{figure}[b] 
\begin{center}\leavevmode
\includegraphics[width=0.6\linewidth]{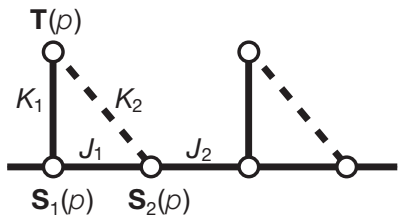}
\end{center}
\caption{ {The quantum spin chain with side chains which 
we study in this paper. 
$\v{S}_1(p)$, $\v{S}_2(p)$ and $\v{T}(p)$ 
are spins in the $p$-th unit cell. 
These magnitudes are $S_1=1$, 
$S_2=\frac{1}{2}$, and $T=\frac{1}{2}$, respectively.}} 
\label{lattice}
\end{figure}

We consider the mixed {Heisenberg spin} chain described by the Hamiltonian
\begin{align}
{\cal H} &=\sum_{p=1}^{N/3} \left[J_1\v{S}_1(p)\v{S}_2(p)+J_2\v{S}_2(p)\v{S}_1(p+1) \right.\nonumber\\
&\left.+ K_1 \v{S}_1(p)\v{T}(p)+K_2 \v{S}_2(p)\v{T}(p)\right].
\label{ham2}
\end{align}
where $\v{S}_2(p)$ and $\v{T}(p)$ are 
 {the spin-$1/2$ operators and $\v{S}_1(p)$ is the spin-1 ones in the $p$-th unit cell, as shown in Fig.~\ref{lattice}.} 
In what follows, we use nondimensional parameters $j=J_2/J_1$, $k=K_1/ {J_1}$ and $r=K_2/(2K_1)$,  {and the unit of $J_1 = 1$}. 
The parameter $r$ characterizes the strength of frustration. 
The total number of spin sites is denoted by $N$, 
and {then} the number of unit cells is $N/3$. 
 
In regime $0 \le r \lesssim 0.5$, where frustration is not strong, 
we have found two types of nonmagnetic ground states 
and have explained them by SCS pictures~\cite{th}. 
When frustration becomes strong, the phase diagram 
drastically changes. 
{In the present paper, w}e will investigate the strongly frustrated case 
{$r \gtrsim 1$} in detail. 

\section{Limiting cases}
\label{limit}

\subsection{Small $j$ regime}
\label{small_j}

For $j=0$, the chain is decoupled into an assembly of 3-spin clusters described by the Hamiltonian
\begin{align}
\HA {(p)}&=\v{S}_1(p)\v{S}_2(p)+ k\v{S}_1(p)\v{T}(p)+2kr\v{S}_2(p)\v{T}(p).
\end{align}
As discussed in  {Ref.}~\cite{th}, the cluster ground state is a singlet state with energy
\begin{align}
E_{\rm s}&=-k-1+\frac{kr}{2} 
\end{align}
 for $k < k_{\rm c}\equiv \frac{2}{r}-1$, and a triplet state with energy
 \begin{align}
E_{\rm t}&=\frac{-(1+k+2kr)-\sqrt{(1+k-4kr)^2+8(k-1)^2}}{4} 
\end{align}
for $k > k_{\rm c}$. Therefore, the ground state of the chain for small $j$ is a gapped local 3-spin singlet phase for  $k < k_{\rm c}$. We call this phase as Gap I phase following  {Ref.}~\cite{th}. 
Because this cluster ground state is gapped, it cannot gain energy within the first order in $j$ even in the presence of $j$.

 { {In} the  {3-spin} triplet ground state for $k > k_{\rm c}$, we define a composed spin with magnitude 1 as 
$\hat{\v{S}}(p)\equiv \v{S}_1(p)+\v{S}_2(p)+\v{T}(p)$ 
to describe low energy phenomena. 
The total Hamiltonian is written as 
$\H = \H^A_0 + \H^{\rm A}_{\rm int}$ with an unperturbed part $\H^A_0 = \sum_p \H^A(p)$ and an interaction part 
\begin{align}
\H^{\rm A}_{\rm int} &=\sum_{p=1}^{N/3} j\v{S}_2(p)\v{S}_1(p+1). 
\end{align}
The perturbation calculation up to the second order in the interaction part $\H^A_{\rm int}$ yields the following effective bilinear-biquadratic Hamiltonian for the composed spin $\hat{\v{S}}(p)$: }
\begin{align}
\H^{\rm A}_{\rm eff}=\sum_{p=1}^{N/3} \left[\JeffA\hat{\v{S}}(p)\hat{\v{S}}(p+1)+\DeffA(\hat{\v{S}}(p)\hat{\v{S}}(p+1))^2\right] , 
\label{heffa}
\end{align}
 {where the effective interaction parameters consist of the first and the second order perturbation terms as 
$J_{\rm eff}^{A}=J_{\rm eff}^{A(1)}+J_{\rm eff}^{A(2)}$ and 
$D_{\rm eff}^{A}=D_{\rm eff}^{A(1)}+D_{\rm eff}^{A(2)}$. }

 {In the first order perturbation, the effective interaction parameters 
coming from $J_{\rm eff}^{A(1)}$ and $D_{\rm eff}^{A(1)}$ are} 
\begin{align}
\JeffA&\simeq -jX(\alpha) , \\
\DeffA&\simeq 0 , 
\end{align}
 {where $X(\alpha)$ is given by} 
\begin{align}
X(\alpha)&=\frac{\alpha}{\sqrt{2}(1+\alpha^2)^2}\left(1-\frac{\alpha}{2\sqrt{2}}\right)\left(1+\frac{\alpha^2}{2}\right) 
\label{X_alpha}
\end{align}
 {with} 
\begin{align}
\alpha&=\frac{2\sqrt{2} {(1-k)}}{\sqrt{(k+1-4kr)^2+8(k-1)^2}-1-k+4kr}. 
\end{align}
For small but finite $j$, the energy of the LMF phase is given by $E_{\rm t}+J^{\rm A}_{\rm eff}$ per  {unit cell} while the energy of the 3-spin singlet (Gap I) phase is given by $E_{\rm s}$ with no first order correction in $j$. 
Therefore, comparing the energies of these two ground states, we find {that} the phase transition between the gapped 3-spin singlet phase and the LMF phase takes place at
\begin{align}
{k=k_{\rm TF}\equiv} k_{\rm c}+\frac{(1-r)(2-2r+r^2)(3-r)j}{2r(2-r)(3-4r+2r^2)}.
\end{align}

The effective coupling $J_{\rm eff}^{A {(1)}}$ vanishes for $k=1$. Therefore, in the neighbourhood of $k=1$, the higher order terms come into play.  Within the second order perturbation with respect to $1-k$ and $j$, the effective coupling constants are 
\begin{align}
\JeffA&
\simeq\frac{j^2}{8(r-1)(1+2r)}-\frac{j(1-k)}{2(2r-1)},\\
\DeffA&
\simeq\frac{j^2}{8(r-1)(4r^2-1)}. 
\end{align}
The effective model (\ref{heffa}) has a variety of phases~\cite{lauchli}. Within the present parameter regime, we find the Haldane phase for $0< \DeffA < \JeffA$ which corresponds to Gap II phase in  {Ref.}~\cite{th}, the QP phase for $0<\JeffA<\DeffA$ and the LMF phase for $\JeffA < 0$. 
However, {in the original Hamiltonian (\ref{ham2}),} 
the LMF phase is limited by the transition to the 3-spin singlet phase at $\JeffA = E_{\rm s}-E_{\rm t}$ {as discussed above}. 
For $\JeffA > E_{\rm s}-E_{\rm t}$, the ground state is the 3-spin singlet (Gap I) phase. 

\begin{figure}
\centerline{\includegraphics[width=70mm]{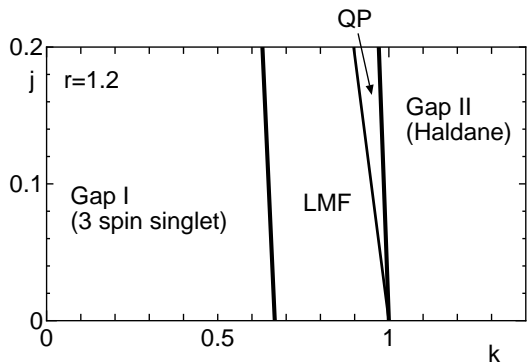}}
\caption{Phase diagram for small $j$ with $r=1.2$.}
\label{phasesmallj}
\end{figure}

Thus the conditions for each phase in terms of the original parameters are summarized as  {follows: 
The ground-state phase of the present model is} 
(i) the 3-spin singlet (Gap I) phase for 
\begin{align}
0 < k < k_{\rm TF}, 
\end{align}
(ii) the LMF phase for
\begin{align}
k_{\rm TF} < k < k_{\rm FQ}\equiv 1-\frac{(2r-1)j}{4(r-1)(2r+1)}, 
\end{align}
(iii) the QP phase for 
\begin{align}
k_{\rm FQ}<{k}<k_{\rm QH}\equiv 1-\frac{j}{2(2r+1)}, 
\end{align}
and (iv) the Haldane (Gap II) phase for 
\begin{align}
k > k_{\rm QH}. 
\end{align}
These phase boundaries are plotted on the $k$-$j$ plane in Fig. \ref{phasesmallj} for $r=1.2$. It should be remarked that the QP phase in the present model is realized without biquadratic interaction in the original Hamiltonian (\ref{ham2}). 

\subsection{Large $j$ Regime}
\label{large_j}

\begin{figure} . 
\centerline{\includegraphics[width=70mm]{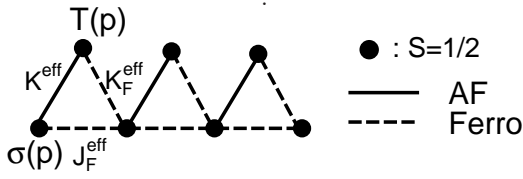}}
\caption{$\Delta$-chain realized in the limit $j >>1$.}
\label{delta}
\end{figure}

In the large $j$ limit, $\v{S}_1(p)$ and {$\v{S}_2(p-1)$} form an effective $S=1/2$ spin $\vg{\sigma}(p)\equiv \v{S}_1(p)+ {\v{S}_2(p-1)}$. The effective Hamiltonian is given by 
\begin{align}
{H} &=\sum_{p=1}^{N/2} (K_1^{\rm eff}\vg{\sigma}(p)\v{T}(p)-J_{\rm F}^{\rm eff}\vg{\sigma}(p)\vg{\sigma}(p+1) \nonumber\\
&- K_{\rm F}^{\rm eff}\vg{\sigma}(p+1)\v{T}(p)),
\label{ham_delta}
\end{align}
which form a $\Delta$-chain structure depicted in Fig. \ref{delta}.
The effective interactions are given by 
\begin{align}
J_{\rm F}^{\rm eff}=\frac{4}{9},\ \ K_{\rm F}^{\rm eff}=\frac{2kr}{3},\ \ K_1^{\rm eff}=\frac{4k}{3},
\end{align}
as  {argued} in  {Ref.}~\cite{th}. 
The detailed analysis of this model is reported in a separate paper ~\cite{delta}.  {Therefore,} 
we only quote  {the results and rewrite them} in terms of the original model (\ref{ham2}).

The ferromagnetic phase of the model (\ref{ham_delta}) corresponds to the LMF phase in the original model (\ref{ham2}). This phase is stable for 
\begin{align}
0 < k < k_{\rm FQ}=\frac{r-2}{3r} 
\end{align}
even in the limit of large $j$.

For $k>k_{\rm FQ}$, the ground state is nonmagnetic. Nevertheless, there are still several different phases. 
For large $r$, 
the model (\ref{ham_delta}) reduces to the $S=1$ bilinear-biquadratic model and the QP phase appear for
\begin{align}
k_{\rm FQ}< k < k_{\rm QH}\equiv \frac{3r-4}{9r},
\end{align}
and Haldane phase appear for $k > k_{\rm QH}$. 

\begin{figure}
\centerline{\includegraphics[width=70mm]{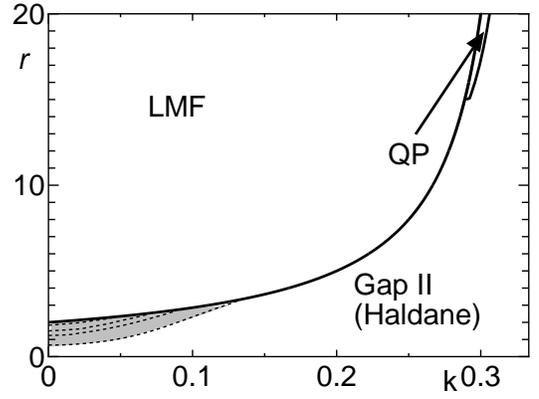}}
\caption{Phase diagram in large $j$ limit. The shaded region is the almost critical nonmagnetic phase.}
\label{phaseovaleff}
\end{figure}

The results of the numerical diagonalization calculation for the model (\ref{ham_delta}) is summarized in Fig. \ref{phaseovaleff}. 
{There are the QP, the Haldane and the LMF phases, 
which we discussed above. 
In addition, numerical results suggests} that there possibly exist 
a narrow  PPF phase between the LMF phase and the QP phase, and almost critical nonmagnetic ground states for small values of $k$. 
The { PPF phase} is so narrow that it cannot be represented in Fig. \ref{phaseovaleff}. 
{We speculate that the almost critical nonmagnetic} phases are spin gap phases with extremely small energy gap with large scale resonating singlet cluster solid structure. 
{Corresponding} phases are also found for finite $j$ as described in the  {next} section.

\subsection{Large  $r$ Regime}
\label{large_r}

 {We examine the case of {$K_2 \gg K_1, J_1$ 
 ($r, kr \gg 1$)} in this subsection.  In the the limit of $r, kr \rightarrow \infty$,} 
the chain is decoupled into an assembly of 3-spin clusters, 
 {each of which} described by the Hamiltonian
\begin{align}
\HB {(p)}&=j\v{S}_1(p+1)\v{S}_2(p)+ 2kr \v{S}_2(p)\v{T}(p).
\end{align}
The eigenvalues of this 3-spin Hamiltonian are given by
\begin{align}
E(S=2)&=\frac{kr}{2}+\frac{j}{2},\\
E(S=1;g)&=\frac{-j-2kr-\sqrt{16k^2r^2-8krj+9j^2}}{4},\\
E(S=1;e)&=\frac{-j-2kr+\sqrt{16k^2r^2-8krj+9j^2}}{4},\\
E(S=0)&=\frac{kr}{2}-j.
\end{align}
The ground states are the triplet states with energy $E(S=1;g)$. 
It should be noted that the lowest excitation energy is of the order of $kr$ even if $j$ is small. Therefore the perturbation calculation from this limit is valid even for small $j$. 
As a result, each cluster has  {an} effective spin-1 degree of freedom  $\tilde{\v{S}}(p)\equiv \v{S}_1(p+1)+\v{S}_2(p)+\v{T}(p)$. 
 {The total Hamiltonian is written as 
$\H = \H^B_0 + \H^B_{\rm int}$ with  {an}  unperturbed part $\H^B_0 = \sum_p \H^B(p)$ and   {an} interaction part }
\begin{align}
\H^{\rm B}_{\rm int}&=\sum_{p=1}^{N/3} (\v{S}_2(p)+k\v{T}(p))\v{S}_1(p).
\end{align}
We can write down the effective Hamiltonian for  $\tilde{\v{S}}(p)$ up to the second order in $\H^{\rm B}_{\rm int}$ in the form,
\begin{align}
\H^{\rm B}_{\rm eff}=\sum_{p=1}^{N/3}\left[ \JeffB\tilde{\v{S}}(p)\tilde{\v{S}}(p+1)+\DeffB(\tilde{\v{S}}(p)\tilde{\v{S}}(p+1))^2\right].
\label{heffb}
\end{align}
Within the first order in $\H^{\rm B}_{\rm int}$, 
 {the effective interaction parameters are given as} 
\begin{align}
\JeffB&\simeq -(X(\alpha)+kX(-\alpha)),\\
\DeffB&\simeq 0 , 
\end{align}
 {where $X(\alpha)$ is given by Eq.~(\ref{X_alpha}) with}
\begin{align}
\alpha&=\frac{j-4kr+\sqrt{16k^2r^2-8krj+9j^2}}{2\sqrt{2}j}. 
\end{align}
The effective exchange constant $\JeffB$ vanishes  {up to the first order in $\H^{\rm B}_{\rm int}$} if $j$ and $k$ satisfy the relation 
\begin{align}
\frac{j}{r}=\frac{8k(k^2-1)}{(3k-1)(3k-5)}.
\end{align}
 {We denote the value of $k$ which satisfies this relation 
by $k_{\rm c}(j/r)$ as a function of $j/r$.} 
The terms {of} $O(r^{-1})$ come into play for $k\simeq k_{\rm c}(j/r)$, {and constitute} the bilinear-biquadratic form (\ref{heffb}) with $\JeffB \sim O(k-k_c,r^{-1})$ and $\DeffB \sim O(r^{-1})$.
 
\begin{figure}
\centerline{\includegraphics[width=70mm]{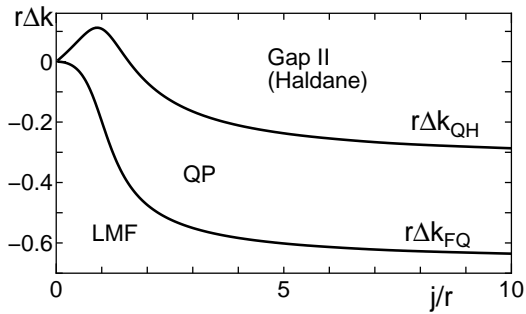}}
\caption{Phase diagram on the $j/r$-$r\Delta k$ plane where $\Delta k=k-k_{\rm c}(j/r)$.}
\label{devi}
\end{figure}

\begin{figure}
\centerline{\includegraphics[width=60mm]{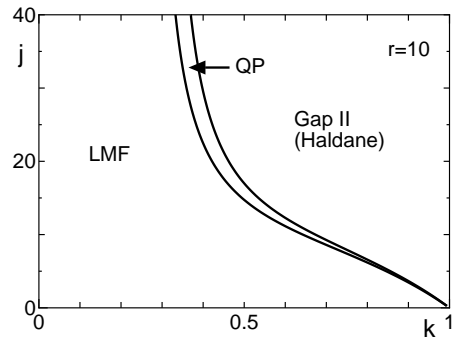}}
\caption{Phase diagram for $r=10$ using the approximation in subsection \ref{large_r}.}
\label{phaseper}
\end{figure}

We do not {explicitly present the second order expression for $\JeffB$ and $\DeffB$, since we numerically carried out} the summation over the intermediate states in the second order perturbation calculation. 
The LMF-QP phase boundary $k_{\rm FQ}$ is determined by setting $\JeffB = 0$. Because the correction terms are {of} $O(r^{-1})$, {deviations} $\Delta k_{\rm FQ}\equiv k_{\rm FQ}-k_{\rm c}(j/r)$ and $\Delta k_{\rm QH}\equiv k_{\rm QH}-k_{\rm c}(j/r)$  {scale} with $1/r$ for fixed $j/r$. 
{Figure \ref{devi} shows the $j/r$-dependence 
 of  $r\Delta k_{\rm FQ}$ and $r\Delta k_{\rm QH}$.} 
{The phase boundaries} for $r=10$ determined by the {present} approximation {are} shown in Fig. \ref{phaseper}.

The calculation in this subsection suggests that the QP phase found in the small-$j$ limit  {(\ref{small_j})} and that in the large-$j$ limit  {(\ref{large_j})} 
{form} a single phase, although it is explicitly demonstrated only in the large $r$ limit. 

\section{Numerical Ground State Phase Diagram}

\begin{figure}
\centerline{\includegraphics[width=70mm]{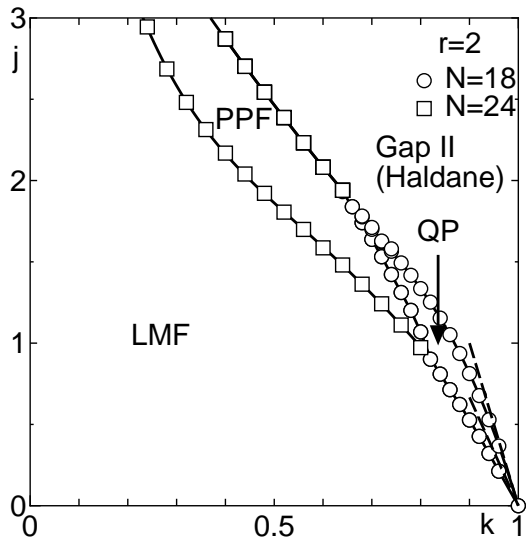}}
\caption{Ground state phase diagram of (\ref{ham2}) for $r=2.0$.  In this and following figures, phase boundaries determined from the numerical diagonalization data for $N=18$ and 24 are shown. The solid lines are guide for eye. The broken lines are the results of the small-$j$ approximation in \ref{small_j}.}
\label{phase20}
\end{figure}
\begin{figure}
\centerline{\includegraphics[width=70mm]{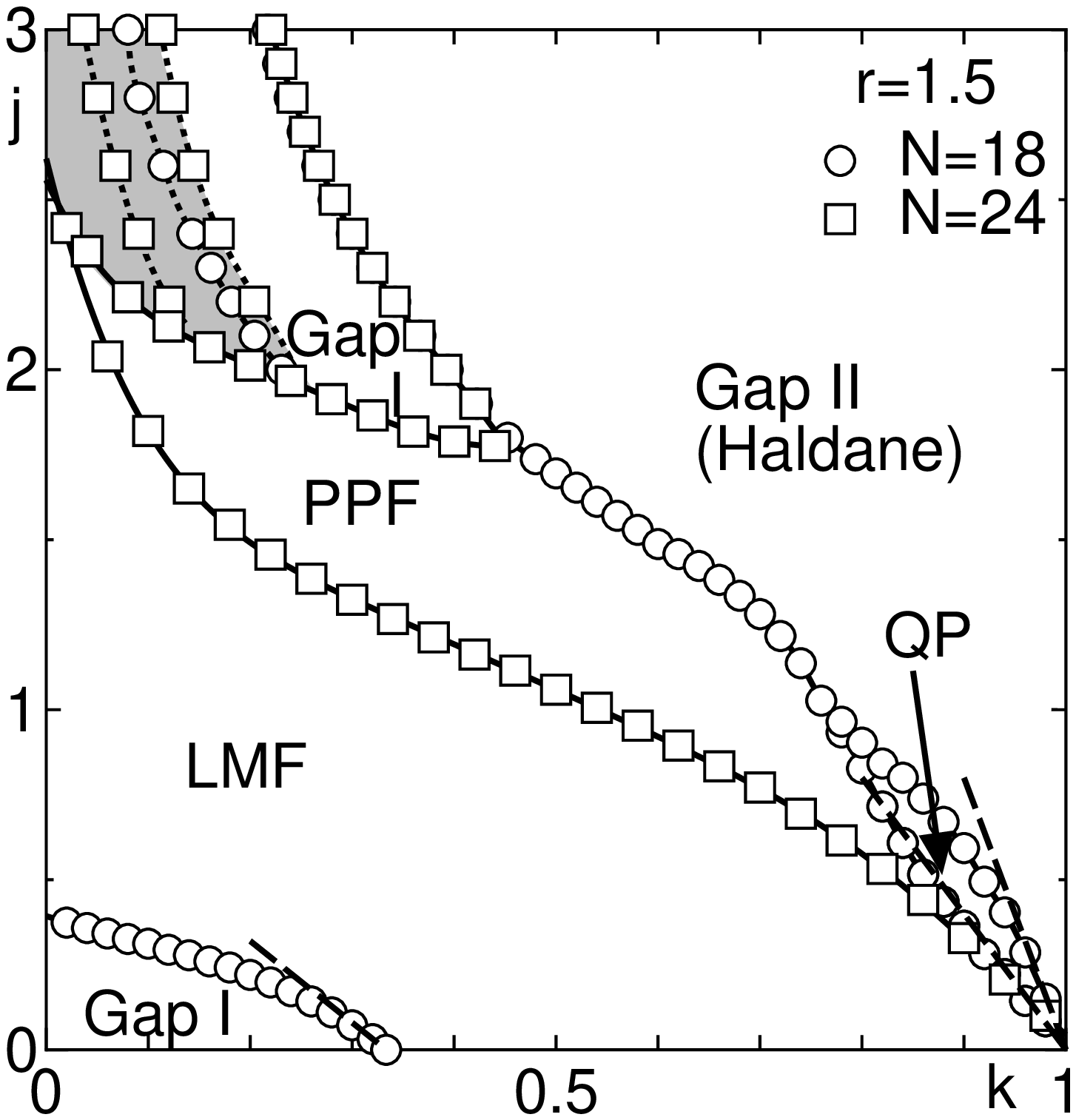}}
\caption{Ground state phase diagram of (\ref{ham2}) for $r=1.5$.  }
\label{phase15}
\end{figure}
\begin{figure}
\centerline{\includegraphics[width=70mm]{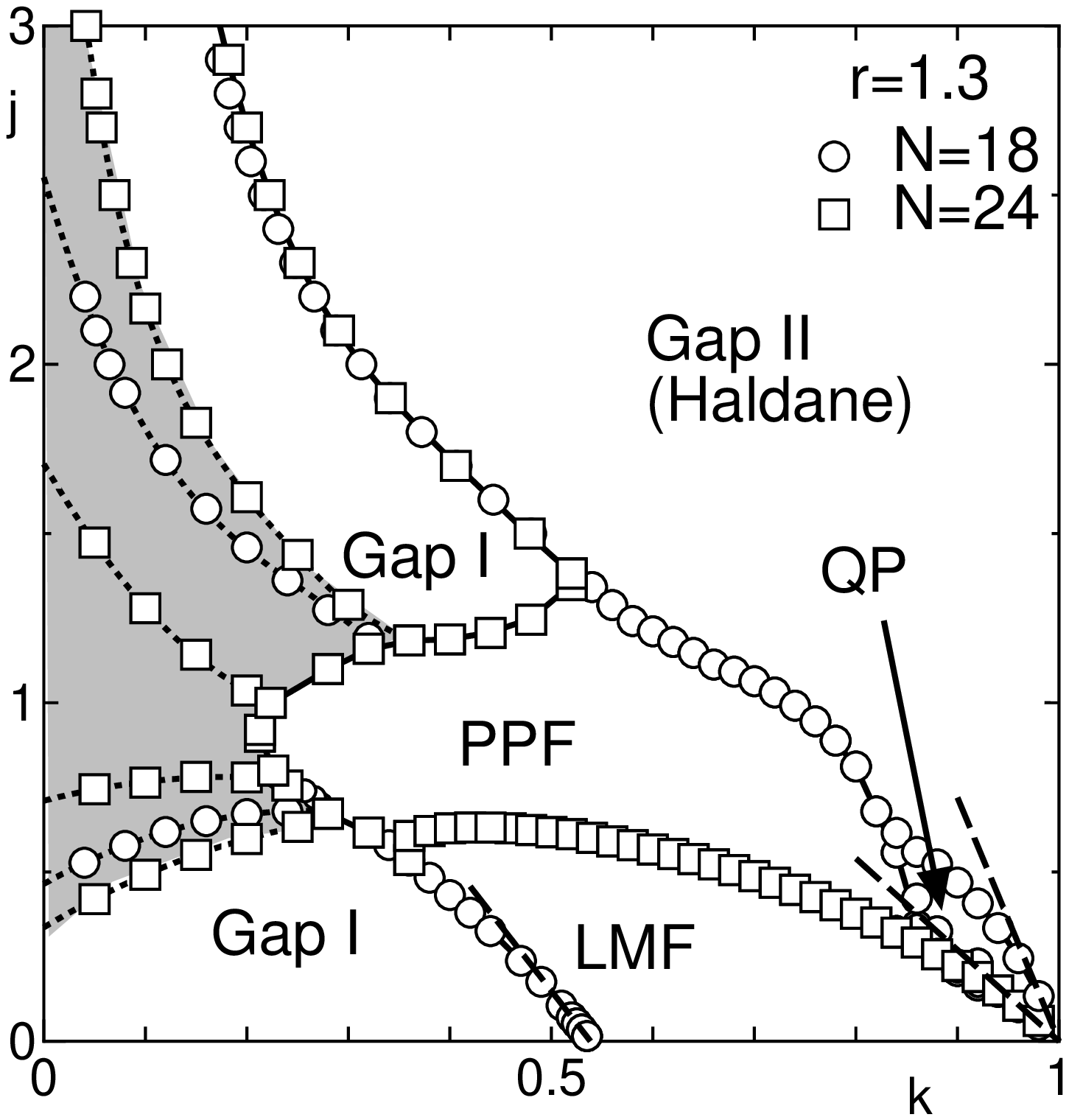}}
\caption{Ground state phase diagram of (\ref{ham2}) for $r=1.3$..
}
\label{phase13}
\end{figure}
\begin{figure}
\centerline{\includegraphics[width=70mm]{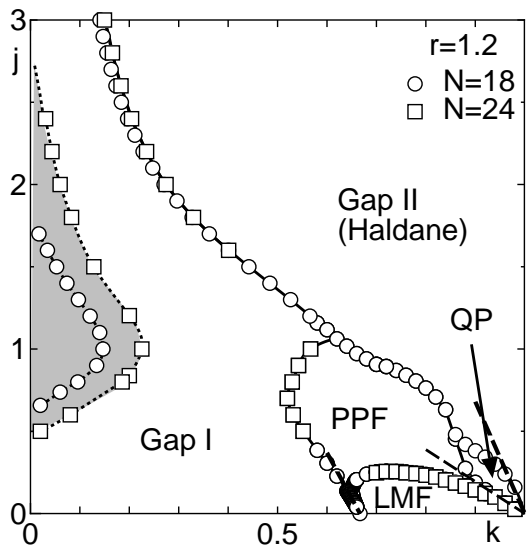}}
\caption{Ground state phase diagram of (\ref{ham2}) for $r=1.2$.}
\label{phase12}
\end{figure}

 For small $r$, there are two types of nonmagnetic ground states and the Gaussian transition occurs between them as described in  {Ref.}~\cite{th}.  {The perturbational approaches in \ref{limit}, however, predict the presence of the LMF and the QP phases  in addition to the conventional spin gap phases.}  

{We start with the case of $r=2$, where the frustration 
is fairly strong. The ground state phase diagram 
is shown in Fig. \ref{phase20}.} 
The phase boundaries calculated {by} using the numerical diagonalization data for $N=18$ and 24 are shown.  Between the Haldane-like Gap II phase and the LMF phase with spontaneous magnetization $\M0=\Msat/2$, there appear a QP phase, which is identified by the lowest excitation with total spin 2,~\cite{fs1,fs2} for {$0.66 \lesssim k \leq 1$} as expected from the perturbational calculation. 
For {$k \gtrsim 0.5$}, only the data for $N=18$ are  {shown} considering the quasi-trimerized nature of the QP phase. In spite of the limited system size, the phase boundary coincides well with the perturbational results for small $j$ as depicted by the broken lines. However, the QP phase vanishes {when $k$ decreases and $j$ increases.}  Instead, there appears a  PPF phase with the spontaneous magnetization of intermediate values between $\Msat/2$ and 0.  The detailed properties of this phase is discussed separately in the next section.

For  $r < 2$, the 3-spin singlet Gap I phase appears for $0< k < k_{\rm c}$. As shown in the phase diagrams  Fig. \ref{phase15} for $r=1.5$,  Fig. \ref{phase13} for $r=1.3$ and  Fig. \ref{phase12} for $r=1.2$, both the QP phase and the  PPF phase shrink to {regions} around $k \sim 1$  {with the decrease}  {of $r$}. 
At $r=1$, the 3-spin singlet phase extends up to $k=1$ for small $j$, and both the QP and the  PPF phases vanish.

 The phase boundary between the Gap I and  the Gap II phases is determined by the twisted boundary method~\cite{kita,kn}. In Fig. \ref{phase15} for $r=1.5$, the Gap I phase consists of two separate regions which have the same parity under the twisted boundary condition. However, these two regions merge with the decrease of $r$ as shown in Fig. \ref{phase12} for $r=1.2$. Therefore we conclude these two regions belong to a single phase. 

\begin{figure} . 
\centerline{\includegraphics[width=80mm]{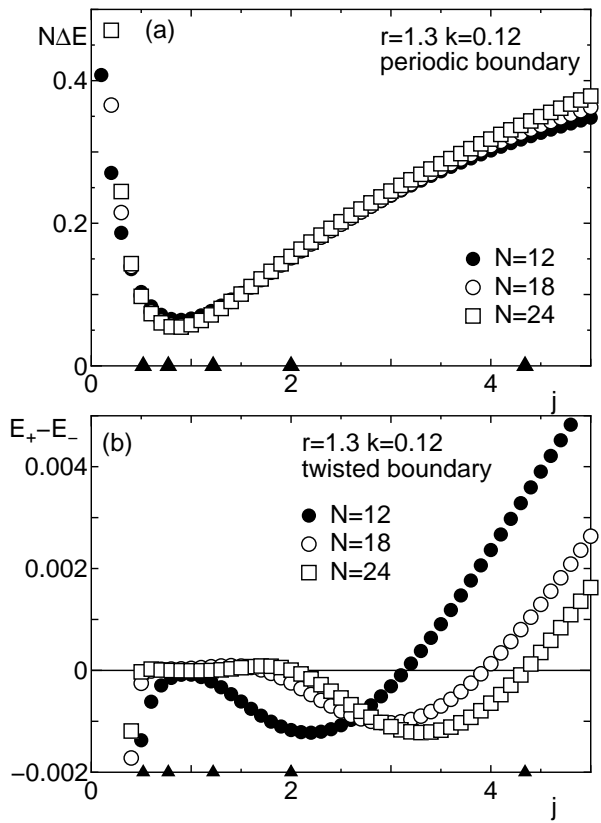}}
\caption{(a) The $j$-dependence of the scaled  {singlet-triplet} energy gap $N\Delta E$ with periodic boundary condition and (b)  the energy difference between the different parity ground states $E_+-E_-$  with twisted boundary condition. The parameters are $r=1.3$, $k=0.12$ and $N=12, 18$ and 24.  The triangles indicate the values of $j$ where the ground state parity changes under the twisted boundary condition for $N=24$.}
\label{gaptwpr}
\end{figure}

 {For small $k$}, we find the region in which the {singlet-triplet} excitation gap $\Delta E$ behaves almost critically.  Figure \ref{gaptwpr}(a) shows the $j$-dependence of the scaled gap $N\Delta E$ for $r=1.3$ and $k=0.12$ for $N=12, 18$ and 24. {Although the boundary of such region cannot be precisely determined, they roughly {correspond to} the shaded regions in  Figs. \ref{phase15}, \ref{phase13} and \ref{phase12}.} 
Applying the twist boundary method,~\cite{kita,kn} 
{we calculate the ground state energies  $E_+$ and $E_-$ with spin inversion parities $+$ and $-$, respectively.  Then} 
we find  that the spin inversion parity of the ground state, 
{or the sign of $E_+ - E_-$,} changes several times {with the variation of $j$} 
as shown in  Fig. \ref{gaptwpr}(b).  
In this region, 
the difference {$E_+ - E_-$} is extremely small (typically less than $O(10^{-3})$ for $N=24$), {and} becomes even smaller with the decrease of $k$. 
This behavior is most {prominent} for $j \sim 1$ as shown in Figs. \ref{phase13} and \ref{phase12}, unless this regime is covered by the ferrimagnetic phase as in Figs. \ref{phase20} and \ref{phase15}.

The critical values of $j$ at which the parity changes are shown by the dotted lines in Figs. \ref{phase15}, \ref{phase13} and \ref{phase12} for each system size. They depend sensitively on the system size. {Due to the limitation of the system size, 
{we cannot} 
conclude whether these lines correspond to {some} phase transitions in the thermodynamic limit. However, for large $j$, these lines are expected to be continuously connected to the similar lines of the effective model (\ref{ham_delta}) in the corresponding regime (the shaded region in  Fig. \ref{phaseovaleff}). The numerically estimated values of the central charge $c$ of the effective model (\ref{ham_delta}) on these lines suggest that they are Gaussian transition lines with $c=1$ among spin gap phases with extremely small gap and large scale singlet clusters~\cite{delta}. Therefore it is likely that these lines in the present model are also similar Gaussian transition lines. However, considering the large ambiguity in the estimation of $c$ in Ref. ~\cite{delta}, other possibilities cannot be ruled out. The elucidation of the nature of the ground state in this regime is left for future studies.}

{\color{red}
\section{Partially Polarized Ferrimagnetic Phase}}
\subsection{Numerical Results}

\begin{figure}
\centerline{\includegraphics[width=70mm]{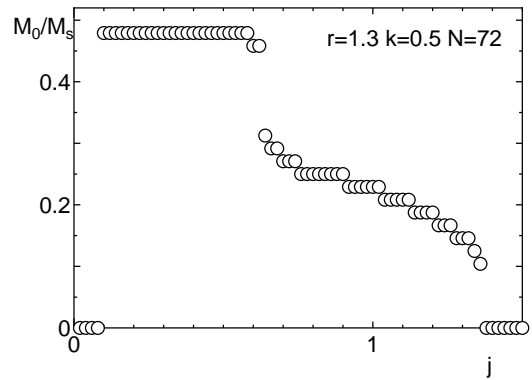}}
\caption{ {The $j$-dependence of the spontaneous magnetization for $k=0.5$ and $r=1.3$. 
The results are calculated in the $N=72$ system by the DMRG method.} 
The magnetization is normalized by the saturated magnetization $\Msat\equiv 2N/3$.}
\label{fig:spont}
\end{figure}

{To clarify properties of the  PPF phase, we calculated the spontaneous magnetization $\M0$ of the ground state by the density matrix renormalization group (DMRG) method. 
In Fig. \ref{fig:spont}, we show $\M0$ as a function of $j$ for $r=1.3$ and $k=0.5$ with $N=72$.} 
For small $j$, the ground state is in the Gap I phase with $\M0=0$. {When $j$ increases}, the LMF phase sets in where $\M0=\Msat/2$. The slight deviation of {$\M0/\Msat$} from 0.5 in this phase is due to the boundary effect inevitable for the open boundary DMRG.   With further increase of $j$, the ground state enters into the  PPF phase where the spontaneous magnetization gradually decreases down to zero. 

\begin{figure}
\centerline{\includegraphics[width=70mm]{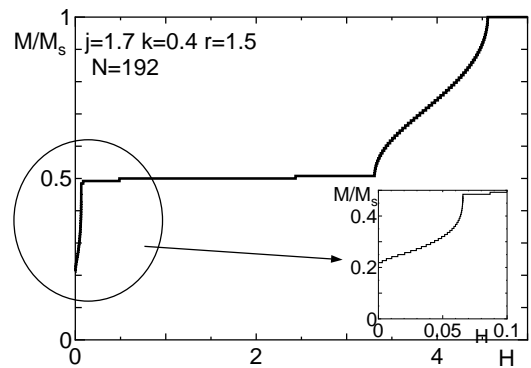}}
\caption{Magnetization curve  in the  PPF phase with $k=0.4, j=1.7$ and $r=1.5$ for $N=192$ calculated by the DMRG method.}
\label{magcurve}
\end{figure}

 Typical magnetization curve calculated by the DMRG is presented in Fig. \ref{magcurve}  for $r=1.5$, $j=1.7$ and $k=0.4$. 
{The} magnetization increases continuously from the zero field value in the  PPF phase in contrast to the LMF phase where {the magnetization is quantized to the zero field value up to a finite critical field~\cite{kura}.} This implies that the magnetic excitation is gapless in the  PPF phase  {as in the previously reported systems~\cite{ss,bkk,ir}.}

\begin{figure}
\centerline{\includegraphics[width=76mm]{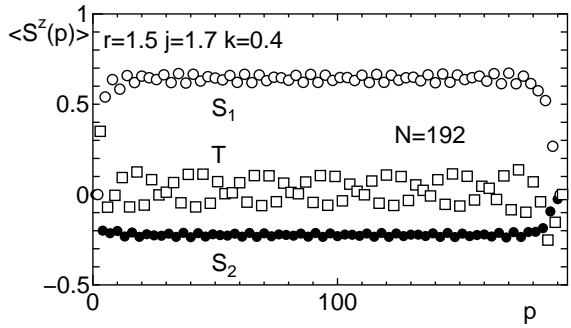}}
\caption{Local magnetization profile in   PPF phase with $k=0.4, j=1.7$ and $r=1.5$ for $N=192$ calculated by the DMRG method.}
\label{magpro}
\end{figure}

  The local magnetization profile $\aver{S^z(p)}$  calculated by the DMRG is plotted against $p$ in Fig. \ref{magpro}  for $r=1.5$, $j=1.7$ and $k=0.4$. In addition to the period 2 oscillation, {an  incommensurate modulation is clearly  observed in the local magnetization.} 

{Similar behaviors are found in other frustration induced  PPF phases} in the spin-1/2 period-3 chain with next nearest 
{neighbor} interaction ~\cite{kh} and in the model  of Ref.~\cite{ym}. We expect these features are common aspects of the frustration induced quantum  PPF phase. 
 
\begin{figure}
\centerline{\includegraphics[width=40mm]{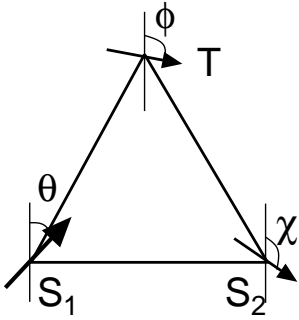}}
\caption{Classical planar spin configuration in a triangle.}
\label{triangle}
\end{figure}
\begin{figure}
\centerline{\includegraphics[width=60mm]{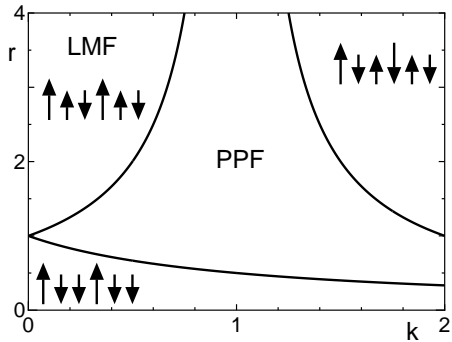}}
\caption{Classical phase diagram on the $k$-$r$ plane.}
\label{cphase}
\end{figure}
\begin{figure} . 
\centerline{\includegraphics[width=60mm]{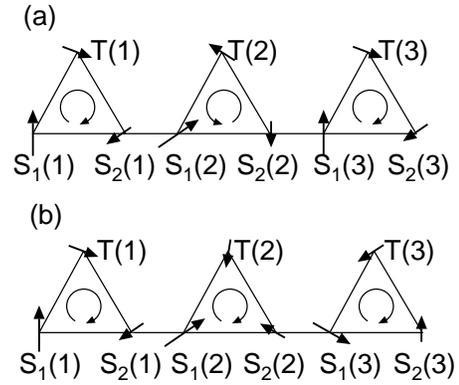}}
\caption{Examples of classical ground state configurations (a) with finite magnetization and (b) with vanishing magnetization. The inner arrows indicate the direction of the rotation of spins along each triangle.}
\label{als}
\end{figure}
\begin{figure}
\centerline{\includegraphics[width=70mm]{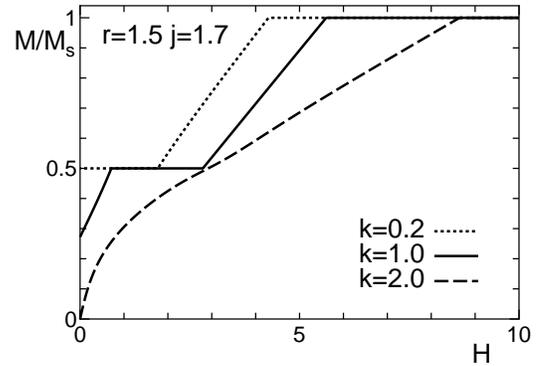}}
\caption{Classical magnetization curves for $k=0.2, 1.0$ and 2.0 with $j=1.7$ and $r=1.5$.}
\label{cmagcurve}
\end{figure}

\subsection{Classical Picture}

To understand the physical picture of the  PPF phase, we consider the classical limit of the Hamiltonian (\ref{ham2}). {Since the $J_2$-bonds are not frustrated, the relative angles between the spins $\v{S}_1(p)$, $\v{S}_2(p)$ and $\v{T}(p)$,  which form a triangle, are not affected by $j$ in the absence of magnetic field. The ground state of the whole chain can be constructed  by arranging the } { triangles so that $\v{S}_1(p)$ and $\v{S}_2(p-1)$ are antiparallel.  The classical ground state energy $E_{\rm G}^{\triangle}$ of a 3-spin cluster consisting of the spins $\v{S}_1(p)$, $\v{S}_2(p)$ and $\v{T}(p)$ is given by
\begin{align}
E_{\rm G}^{\triangle}(\phi, \theta, \chi)&=\frac{k}{2}\cos(\phi-\theta)+\frac{kr}{2}\cos(\phi-\chi)+\frac{1}{2}\cos(\chi-\theta) , 
\end{align}}
{assuming a planar configuration. We confine ourselves to the planar configuration, because  nonplanar configurations have higher energy. The angles $\theta$, $\phi$ and $\chi$ denote the polar angle of $\v{S}_1(p)$, $\v{T}(p)$ and $\v{S}_2(p)$ measured from $z$-axis as shown in Fig. \ref{triangle}, respectively.   By minimizing $E_{\rm G}^{\triangle}$, we find that a stable noncollinear configuration within the triangle is realized in the region
\begin{align}
\left|\frac{r-1}{r}\right| < k < \frac{r+1}{r} 
\label{confc}
\end{align}
}
{which is indicated as  PPF in   Fig.~\ref{cphase}. Outside this region, three different collinear configurations are flavored depending on the values of $k$ and $r$ as
\begin{align}
\ket{S_1^z(p)T^z(p)S_2^z(p)}=\left\{
\begin{array}{cc}
\ket{\Uparrow\downarrow\uparrow} & k > \dfrac{r+1}{r},\\
\ket{\Uparrow\uparrow\downarrow} & 0 < k < \dfrac{r-1}{r}, \\
\ket{\Uparrow\downarrow\downarrow} & 0 < k < \dfrac{1-r}{r}.
\end{array}
\right.
\end{align}
If these are regularly arranged along the chain keeping the spins $\v{S}_1(p)$ and $\v{S}_2(p-1)$ antiparallel, the LMF phase and two kinds of N\'eel phases are realized as indicated in Fig.~\ref{cphase}. }

In the region (\ref{confc}), however, the energy is invariant under the simultaneous rotation of two of the spins (say $\v{S}_2(p)$ and $\v{T}(p)$) around the remaining one (say $\v{S}_1(p)$). 
Therefore the ground state  of the whole chain has a {macroscopic degeneracy.  Among these highly degenerate ground states, the states with various values of magnetization are included. For example, the state depicted in Fig. \ref{als}(a) has finite magnetization, while the other one depicted in Fig. \ref{als}(b) has a spiral structure and  has no net magnetization. }

{In the presence of magnetic field, the degeneracy is lifted and one of the ground states which has the  largest magnetization is selected by an infinitesimal magnetic field.  In this state, the spin configuration should be restricted in a single plane, because the tilt out of the plane reduces the net magnetization. 
{Then each triangle in the spin chain takes 
one of twofold degenerate spin configurations with 
clockwise and counterclockwise spin rotations.} 
These two configurations are distinguished by chirality. 
In a uniform array of triangles with the same chirality, the net magnetization tends to be averaged out. Therefore, the magnetization per triangle decreases with {increasing} length of the array. To maximize the magnetization, an array of the triangles with alternating chirality depicted in Fig. \ref{als}(a) is most favorable. This also explains the period 2 oscillation observed in Fig. \ref{magpro}. With the increase of the magnetic field, the spin orientation gradually changes to the direction of the magnetic field}. 

{The ground state in a finite magnetic field $H$ is obtained by numerically minimizing the classical energy
\begin{align}
E_{\rm G}(H)&=\sum_{p=1}^{N/3}\left[\frac{k}{2}\cos(\phi(p)-\theta(p))+\frac{kr}{2}\cos(\phi(p)-\chi(p))\right.\\
&+\frac{1}{2}\cos(\chi(p)-\theta(p))+\frac{j}{2}\cos(\chi(p-1)-\theta(p))\\
&\left.-H(\cos\theta(p)+\frac{1}{2}\cos\phi(p)+\frac{1}{2}\cos\chi(p))\right].
\end{align}
where the magnetic field is in $z$-direction. Considering the argument in the preceding paragraph, we assume a period 2 structure in the minimization procedure. Typical magnetization curves are shown in Fig. \ref{cmagcurve} for $r=1.5$ and $j=1.7$ with $k=0.2, 1.0$ and 2.0. 
At first sight, these three curves correspond to the LMF phase ($k=0.2$),  the  PPF phase ($k=1.0$) and a nonmagnetic phase ($k=2.0$). 
However, the classical phase including the point of $k=2.0$  correspond to a  N\'eel ordered state 
$\ket{S_1^z(1)T^z(1)S_2^z(1)S_1^z(2)T^z(2)S_2^z(2) \cdots}$ = 
$\ket{\Uparrow\downarrow\uparrow\Downarrow\uparrow\downarrow \cdots}$ as discussed above. 
Nevertheless, we may regard this N\'eel phase as a classical counterpart of the Gap II (Haldane) phase, 
since the short range antiferromagnetic correlation between the total spins of the 3-spin clusters is common in the N\'eel state and 
in the Haldane-like state, and the latter is more stable in the quantum case}.  {Similarly, in the classical  PPF ground state has noncollinear spin structure with broken $U(1)$ symmetry around the $z$-axis, while in the quantum case, the $U(1)$ symmetry is restored due to quantum fluctuation.} 
It should be also noted that another ordered ground state with structure $\ket{\Uparrow\downarrow\downarrow\Uparrow\downarrow\downarrow...}$ corresponding to the classical counterpart of the 3-spin singlet phase appears for small $r$ as shown in Fig. \ref{cphase}.

Thus, all the phases in the {quantum} model (\ref{ham2}) have the classical counterparts except for the QP phase which is of essentially quantum origin. Therefore the classical picture appears to be satisfactory at least qualitatively.  The parameter regime for each phase is, however, largely different from  {that of} the original quantum model (\ref{ham2}). For example, the condition for the classical noncollinear spin configuration (\ref{confc}) is independent of $j$ and does not cover the numerically obtained region of the  PPF phase. In addition, the incommensurate modulation of the spin profile is not explained in the classical model. These features are essentially quantum effect which is beyond the classical interpretation.

\section{Summary and Discussion}

A mixed spin chain with frustrated side chains is investigated, 
when the frustration is strong. 
{Not only the LMF phase but also the  PPF phase appears between the nonmagnetic phase and the LMF phase. 
The  PPF phase has continuously varying spontaneous magnetization, which is not a simple fraction of the saturated magnetization.} 
The local magnetization profile has {an} incommensurate structure in the  PPF phase. These features are common with other examples of  PPF phases induced by frustration. 
{Classical interpretation of the  PPF phase is also presented. It is pointed out that an infinitesimal magnetic field selects the  PPF state in an appropriate parameter region.} 
{For the quantum model, however, the  PPF state is realized 
in the absence of magnetic field. 
This {suggests} 
that quantum fluctuation 
 selects one of the classical ground states.}

The presence of QP phase is demonstrated by the perturbational  {calculation} as well as the numerical method. 
Although the  {present} perturbational  {calculation} is  {carried out} for  {our} specific model (\ref{ham2}), 
the derivation of the effective spin-1 bilinear-biquadratic chain is  {quite general and is applicable to models which have} an effective spin-1 degree of freedom in each unit cell. 
 {The QP phase appears around} the point where the first order effective coupling vanishes due to frustration. 
{Hence the QP phase is expected to be commonly found 
in a wide variety of frustrated quantum spin chains.} 

For {$r \gtrsim 1$} and $j \sim 1$ and $k \ll 1$, 
{the main chain couples only weakly with side spins 
but the frustration is strong.} 
{In this case,} there appears an almost critical nonmagnetic ground state in which the spin inversion parity  under the twisted boundary condition changes many times with the variation of parameters. 
{Considering the continuity to the similar ground state in the effective model for large $j$, it is likely that a series of Gaussian transitions take place among gapped phases  with an extremely small gap.} 

{Although the nature of 
this almost critical ground state remains unresolved, we may speculate its physical origin in the following way: In the absence of side spins  $\v{T}(p)$, the ground state of the main chain is ferrimagnetically ordered. For small $k$, side spins are coupled to this ferrimagnetic moment antiferromagnetically via $K_1$ bond and ferromagnetically via $K_2$ bonds. For $r \sim 1$, the effective coupling is even weakened due to frustration, and  for  {$j = 1$}, the ferrimagnetic state of the main chain has no local valence bond structure. Therefore the effective coupling among the side spins, which is mediated by the fluctuation in the main chain, would be very long ranged for $j \sim 1$. This implies that the resultant nonmagnetic state should have a highly nonlocal character. Thus we may speculate that the ground state has an extremely small energy gap and large scale singlet clusters. A similar ground state with extremely small gap is known in a $S=1/2$ zigzag chain with ferromagnetic nearest neighbor interaction and antiferromagnetic next nearest neighbor interaction~\cite{iq}. However we do not find an explicit mapping of the present model onto the field theory of  {Ref.}~\cite{iq}.}

\section*{ACKNOWLEDGMENTS}

 The numerical diagonalization program is based on the package TITPACK ver.2 coded by H. Nishimori.  The numerical computation in this work has been carried out using the facilities of the Supercomputer Center, Institute for Solid State Physics, University of Tokyo and Supercomputing Division, Information Technology Center, University of Tokyo.  This work is {partly} supported by a Grant-in-Aid for Scientific Research  on Priority Areas, "Novel States of Matter Induced by Frustration", from the Ministry of Education, Science, Sports and Culture of Japan, 
 {and  
Fund for Project Research in Toyota Technological Institute}.


\begin{thebibliography}{10}
\bibitem{ncf}  {In Refs. ~\cite{ym,kh}, this phase is called 'noncollinear ferrimagnetic phase' which is based on the classical picture. However, this term gives the impression that the U(1) symmetry around the $z$-axis is spontaneously broken in this phase. Of course, this is not the case in the present one-dimensional model. Therefore we use the term 'partially polarized ferrimagnetism' following ~\cite{ir} to avoid confusion.}
\end{thebibliography}
\end{document}